\documentclass[aps,prb,twocolumn,amsmath,amssymb]{revtex4}
\usepackage{graphicx}
\usepackage{bm}
\usepackage{multirow}
\usepackage{times}
\usepackage{color}
\usepackage{wasysym}

\begin{document}
\title{Phase diagram of the 1/3-filled extended Hubbard model on the Kagome lattice}
\author{Karim Ferhat and Arnaud Ralko}
\affiliation{
Institut N\'eel, UPR2940, Universit\'e Grenoble Alpes et CNRS, Grenoble,
FR-38042 France
} 
\date{\today}
\begin{abstract}
We study the phase diagram of the extended Hubbard model on the kagome lattice
at 1/3 filling. By combining a configuration interaction approach to an
unrestricted Hartree-Fock, we construct an effective Hamiltonian which takes
the correlations back on top of the mean-field solution.
We obtain a rich phase diagram with, in particular, the presence of two
original phases.
The first one consists of local polarized droplets of metal standing on the hexagons
of the lattice, and an enlarged kagome charge order, inversely polarized, on
the remaining sites.
The second, obeying a local {\it ice-rule} type constraint on the triangles of
the kagome lattice, is driven by an antiferromagnetically coupling of spins and
is constituted of disconnected six-spin singlet rings.
The nature and stability of these phases at large interactions are studied via
trial wave functions and perturbation theory.  \end{abstract}
\maketitle

\section{Introduction} 

Lattice systems with geometrical frustration are at the center of intense
theoretical and experimental research activity of current condensed matter,
thanks to the extremely rich exotic phases encountered. One of the most
important features of frustrated systems is the presence, in the classical
limit, of highly degenerate ground states. These states, under the presence of
quantum fluctuations, yield unconventional phases generic either in fermionic
or bosonic systems such as spin liquids\cite{Misguich,Hermele},  valence bond
physics\cite{Rokhsar,Moessner}, pinball phases\cite{Pinball1,Pinball2},
fractionalized defects\cite{Ruegg2}, and even
supersolids\cite{Ralko0,Trousselet}. Hence, frustration is at the crossroads of
many disciplines, ranging from quantum antiferromagnets to optical lattices,
via itinerant electrons and bosonic systems.

Intensively studied in quantum magnetism for exotic phases and spin liquid
ground states, the role of the frustration is more and more considered in
fermionic systems \cite{Pinball1, Pinball2,Ruegg2,Wen,Pollmann}. Among the
typical two-dimensional lattices possessing a geometrical frustration, one can
mention the triangular, checkerboard\cite{Ralko0,Wessel},
Cairo-pentagonal\cite{Ralko3,Ioanis}, and kagome \cite{Ralko1,
Ralko2, Isakov01, Indergrand02}.
The latter is the most frustrated two-dimensional lattice and, as a function of
the electron filling, possesses several induced local constraint phases.
At half filling, under strong Hubbard repulsion, several possible ground states
have been proposed for the Heisenberg model\cite{book}, ranging from valence bond
solids\cite{Marston, Balents, Singh,Evenbly,Poilblanc} to various spin liquids such as a
{\it U(1)} algebraic\cite{Ran}, a triplet-gapped\cite{Jiang01} and a ${\mathbb Z}_2$
topological state with anyonic excitations\cite{Yan,Depenbrock,Jiang02}.
At intermediate interaction and for various fillings, original electronic
features have recently been pointed out, such as peculiar Mott
transition\cite{Ohashi,Laeuchli}, anomalous quantum Hall
effects\cite{Ohgushi,LeHur}, Fermi surface instabilities \cite{Kiesel},
competing orders\cite{Wang}, and even anyonic states \cite{Ludwig}, to name a few.

Surprisingly, at lower filling, only a few works have been done so far. In particular,
the $1/3$-filled system takes a special role since, in the non-interacting
limit, the Fermi sea is at the {\it Dirac cones} at the corner of the Brillouin
zone (the {\it K} point) and the ground state (GS) is a semi-metal (SM).
With interactions, let us cite studies on topological phases
\cite{Wen}, Mott-insulator transitions for spinless fermions \cite{Nishimoto}
or fractionalization of defects with an extra bow-tie term \cite{Ruegg2}.
Under the presence of the Hubbard interaction, spin charge density waves with
antiferromagnetic spin stripes are stabilized\cite{Wen}, but this phase was
enforced to preserve the translations.
With nearest-neighbor interaction, a local constraint emerges, often called the {\it
ice-rule}, very reminiscent of those observed in spin-ices.  Thus, the
ground state manifold is macroscopically degenerate with a finite entropy at
zero temperature. This has to be connected with the physics of hard-core bosons
which arises in corner-sharing frustrating unit lattices such as the
checkerboard \cite{Ralko0,Indergrand02}, the pyrochlore and the Cairo-pentagonal lattices
\cite{Ralko3, Ioanis} for example.
\begin{figure}[ht]
\includegraphics[width=0.45\textwidth,clip]{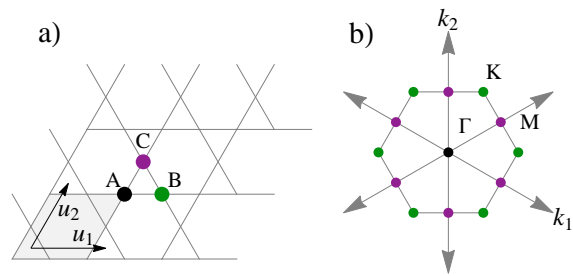}
\caption{(Color online).  a) Example of a $3\times3$ kagome lattice with 
three sites per unit cell dubbed $A$, $B$ and $C$. b) The Brillouin zone with ${\bf
k}_1$ and  ${\bf k}_2$ being the reciprocal vectors of  ${\bf u}_1$ and ${\bf u}_2$
and the elementary translational vectors in real space defined in (a). The highest
symmetry points $\Gamma$; $M$ and $K$ are depicted.
\label{lattice}
}
\end{figure}

On such a manifold, the quantum fluctuations select a resonating plaquette
phase by an order-to-disorder mechanism through a perturbation theory of the
third order similar to that of Resonating Valence Bond (RVB) physics, and the local ice-rule constraint
enforces the system to have exactly one doubly occupied site per triangle
\cite{Ruegg2}.
Despite the presence of all the ingredients --frustration, local constraint,
ground state degeneracy, etc-- needed to observe a very rich and exotic
physics, the phase diagram of this model is still lacking in the literature,
even at the mean-field level.

In this paper, we provide a theoretical study of the phase diagram of the
extended Hubbard model at 1/3 filling at low and intermediate on-site and
nearest neighbor interactions.
In Sec. \ref{methods} are presented the model and methods, as well as the
physical quantities used to determine the phases.  In particular, we briefly
introduce the configuration interaction
(CI) method \cite{Malli,Friedman,Louis2,Li,Bunge,Sambataro} that we combine to an
Unrestricted Hartree-Fock in order to partially bring back the correlations
lost in the mean-field decoupling.
Section \ref{overview} gives an overview of the thermodynamic limit phase
diagram and a brief description of the encountered phases, emphasizing two
original ones,  the pinned metal droplet phase (PMD) and a spin charge density wave
(SCDW).
The identification of the phases, the extrapolation to the thermodynamic limit,
and the stability of the phases upon the correlations by the CI are discussed
in Sec. \ref{TL}.
Finally, the structure of the phases beyond the UHF and CI is discussed in
Sec. \ref{Variational}. We present a variational theory that helps in the
understanding of the mechanism stabilizing the two peculiar phases, one being
the presence of local polarized droplets of metal isolated by pins, and the
other consisting of disconnected hexagons of six-spin singlets.

\section{Model and methods}
\label{methods}

The model considered in this paper is the extended Hubbard model on the kagome
lattice:
\begin{eqnarray} {\cal H} = - t \sum_{\langle i,j \rangle}
( \xi_{i}^\dagger \xi_{j} + h.c )
+ U \sum_i n_{i,\uparrow} n_{i,\downarrow} +
V \sum_{\langle i,j \rangle} n_i n_j \label{ham}
\end{eqnarray}
where $\xi_i^\dagger = (c_{i,\uparrow}^\dagger , c_{i,\downarrow}^\dagger)$ is a
spinor of fermion (electron) creation operators at site $i$,
$n_{i,\sigma} = c_{i,\sigma}^\dagger c_{i,\sigma}$ is the number of electrons
with spin $\sigma$ and  $n_i =  n_{i,\uparrow} + n_{i,\downarrow}$ is the total
density.
$U$ is the on-site Hubbard term, $V$ is the nearest neighbor Coulomb repulsion and
$t >0$ is the hopping term.  Throughout the paper, the lowest band is completely
filled, corresponding to the 1/3 filling.
As reported in the literature, at small interactions $U$ and $V$, the system is
a semi-metal (SM), while at large $V$ (keeping $U$ small) a resonating plaquette charge
density wave (CDW) is stabilized \cite{Wen, Ruegg2, Nishimoto, Wen66,
Pollmann}.
Indeed, in this large $V$ limit, the system is governed by a local constraint
enforcing each triangle of the lattice\cite{Wen67} to have exactly only one
doubly occupied site and the others empty. This ensures the ground state, at
the electrostatic limit ($t=0$), to be macroscopically degenerate. Then, a ring
exchange of particles on a hexagon of order $t^3/V^2$ lifts this degeneracy and
selects the CDW. How those two phases behave upon intermediate values of $U$ is
one of the questions addressed in this work.

The kagome lattice with periodic boundary conditions is defined by its linear
size $L$, and has three sites per unit cell, called $A$, $B$ and $C$ in
Fig.\ref{lattice}. The number of sites is given by $N = 3 L^2$.
Most of the computations of the ground states performed in this paper have been
done by (i) using an Unrestricted Hartree-Fock (UHF) and (ii) partially
bringing back on the top of this mean-field solution the correlations by using
a  Configuration Interaction on single and double excitations (SDCI) of the UHF
Slater determinant\cite{Malli,Friedman,Louis2,Li,Bunge,Sambataro}.
As a reminder, in the standard UHF approach, the interaction terms $n_{i,\sigma} n_{j,\sigma'}$ are
decoupled as a Hartree term  
$\langle n_{i,\sigma}\rangle
n_{j,\sigma'} +  n_{i,\sigma} \langle n_{j,\sigma'} \rangle -  \langle
n_{i,\sigma} \rangle \langle n_{j,\sigma'} \rangle $
and a Fock term $
\langle c_{i,\sigma}^\dagger
c_{j,\sigma'} \rangle c_{j,\sigma'}^\dagger c_{i,\sigma} +
c_{i,\sigma}^\dagger c_{j,\sigma'} \langle c_{j,\sigma'}^\dagger c_{i,\sigma}
\rangle 
-  
\langle c_{i,\sigma}^\dagger c_{j,\sigma'} \rangle \langle
c_{j,\sigma'}^\dagger c_{i,\sigma} \rangle $.
The sets $ \langle n_{i,\sigma} \rangle $ and $\langle c_{i,\sigma}^\dagger
c_{j,\sigma'} \rangle $ , for $i$ and $j\in [1, N]$, are computed
self-consistently, as well as the single electron basis wave functions.
The configuration interaction (CI) method \cite{Malli,Friedman,Louis2,Li,Bunge,Sambataro},
then consists in  constructing an effective
Hamiltonian $H^{\textrm{eff}}_{ij}$ by projecting the microscopic Hamiltonian
${\cal H}$ in a subspace made of Slater determinants $\{ \phi_i \}$,
constructed in this single electron basis obtained by the UHF.
If all Slater determinants are kept in the basis, then the full Hilbert space
is spanned and $H^{\textrm{eff}}_{ij}$ is nothing else but ${\cal H}$. This is
usually referred to as the full CI (FCI) in the literature\cite{Malli}.  Obviously, in this
case, the limitations are the same as those for the exact diagonalizations (EDs),
the Hilbert space is increasing exponentially, and only small cluster sizes are
available.
An advantage to working with the Slater basis however, is that a simple energy
criterion can be exploited.  Indeed, at small and intermediate values of $V$
and $U$, the correlations are expected to be well taken into account by single
and double excitations from the Fermi sea.  Hence, it is reasonable to truncate
the basis to those processes and to restrict our effective Hamiltonian to this
subspace. This is the so called single-double CI (SDCI). Note that this has
been intensively studied in quantum chemistry where it has been shown that, for
certain molecules, up to 80\% of the correlation energy was restored. Here, in
addition to our results, we show in Sec.~\ref{TL}-B an improvement of about 60\%
for certain phases.

In order to characterize the phase diagram, we define a set of order
parameters.
The structure factor $M({\bf q}) = \sqrt{\langle \psi_0 | n(-{\bf q})n({\bf q})
| \psi_0 \rangle / N}$, where $n({\bf q}) = n_\uparrow({\bf q}) +
n_\downarrow({\bf q})$ indicates spontaneous translational symmetry breaking of
the ground state.
Since at very large Coulomb repulsion $V$ the system is governed by the local
{\it ice-rule} constraints, we define the local triangle operators $\xi^{\pm}
= \frac {6}{N} \sum_t w^0 {\rho}^{\pm}_{A} + w^1 {\rho}^{\pm}_{B} + w^2
{\rho}^{\pm}_{C}  $ with ${\rho}^{\pm}_i = n_{i \uparrow} \pm n_{i \downarrow}$
and $w = e^{i 2 \pi / 3}$. $\xi^+ = 1$ for a phase obeying an ice-rule of two
particles per triangle with exactly one doubly occupied site, and $\xi^- =1$ if
the ice-rule of the type of two particles with one empty site is realized. Both
order parameters are zero in the homogeneous phase, and finite otherwise.
To study the internal spin ordered phases, we define the operator
${\rho}^{-}_{A} {\rho}^{-}_{B} {\rho}^{-}_{C}$, which is finite if there are no
vacant sites.

\section{Overview of the phase diagram} 
\label{overview}
\begin{figure}[ht] 
\includegraphics[width=0.45\textwidth,clip]{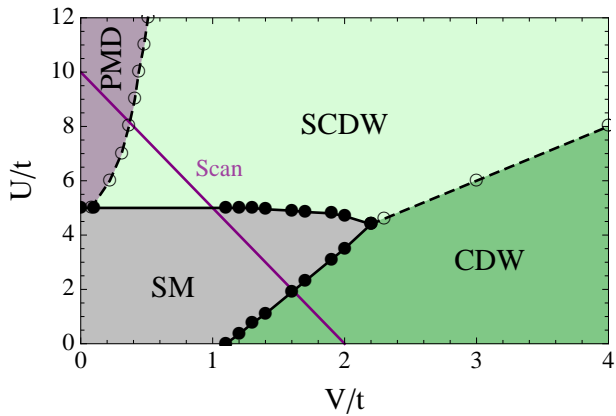} 
\caption{(Color online).  Phase diagram in the $(V,U)$ plane. All points are
extrapolations at the thermodynamic limit obtained by size scalings. The four
phases are the semi-metal (SM), the resonating charge density wave (CDW), the
spin charge density wave (SCDW), and the pinned metal droplets (PMDs).
Dashed (continous) lines and empty (filled) circles correspond to first-order
(second-order) order transitions.
Results for the scan line (purple) are detailed in Fig.\ref{scan}.
\label{PhaseDiag}
}
\end{figure}

\begin{figure}[ht] 
\includegraphics[width=0.42\textwidth,clip]{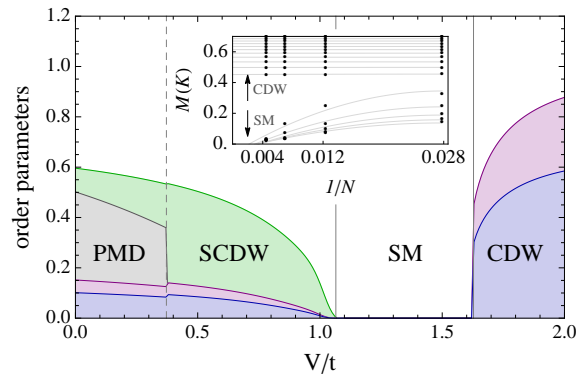} 
\caption{
\label{scan}
(Color online). Behavior at the thermodynamic limit of the order
parameters defined in the text along the scan line depicted in
Fig.\ref{PhaseDiag} corresponding to $U = 10 - 5 V$:
 (blue) $M(K)$, (gray) $\rho^-_A \rho^-_B \rho^-_C $, (green) $\xi^-$, and (red)
$\xi^+$.
Four distinct domains are
observed, separated by continuous (second order) and dashed (first order)
lines. Inset: size scaling of the CDW order parameter close to the transition
with the SM; a clear change of behavior indicates the location of the phase transition.
}
\end{figure}

The phase diagram of the 1/3-filled extended Hubbard model on the kagome
lattice extrapolated to the thermodynamic limit from the UHF results up to $3
\times 36 \times 36$ site clusters is depicted in Fig.\ref{PhaseDiag}. 
Four phases have been obtained in the $(V,U)$ plane, the semi-metal (SM), a
charge density wave (CDW), a spin charge density wave (SCDW), and a pinned metal droplet phase
(PMD). As already mentioned, the first two phases have already been
reported elsewhere\cite{Nishimoto,Wen66,Wen67}. For the CDW, a perturbative
effective model of order $t^3/V^2$ maps the present model onto a quantum dimer
model on the hexagonal lattice\cite{Nishimoto, Wen66, Wen67, Pollmann}. Indeed, in
the classical limit $t = 0$ and $V \neq 0$, the ground state is macroscopically
degenerate, and the configurations that minimize the energy obey a rule of {\it
two electrons per triangle with one doubly occupied site}.  A finite $t$ lifts
this degeneracy and in the limit $t \ll V$ the system is effectively described
by a hardcore quantum  dimer model (QDM) on the honeycomb lattice \cite{Pollmann,Nishimoto,Wen66}
whose ground state consists of resonating plaquettes. The mean-field treatment
cannot capture the resonating plaquettes of the QDM, but its solution can be
thought of as its electrostatic counterpart\cite{Wen}. This can be seen in the UHF
local densities in Fig.\ref{snapshots}. A second order transition between the
SM and the CDW is obtained.
To the best of our knowledge, the two other phases reported in this work,
namely the PMD and SCDW phases, have not been reported so far\cite{Pollmann2}. 
Starting from the CDW phase in the large $V$ regime, by increasing $U$ with $U
> 2 V$, it is more favorable for the system to forbid double occupancies. This
 is happening through a first order phase transition. In contrast to the
recent paper of Wen {\it et al.} [\onlinecite{Wen}], since we do not restrict
our phases to preserve translational symmetries, we obtain a $\sqrt{3} \times
\sqrt{3}$ unit-cell phase with six-spin antiferromagnetic hexagonal rings. All
spin-{\it singlet} rings are disconnected by empty sites, as depicted in
Fig.\ref{snapshots}-a (see the text for more details). We call this phase the
spin charge density wave (SCDW). Its phase transition with the SM is of second
order.
Eventually, at lower $V$ interaction, the antiferromagnetic rings are not
favored anymore, and the system prefers to separate up and down spins via
another first order transition.
One spin sector crystallizes on one third of the sites in order to form an
enlarged kagome structure, the other sector forming disconnected half-filled
hexagonal metal rings on the remaining sites.
This phase, called the {\it pinned metal droplets} (PMDs) phase, since the hexagons
(metal droplets) are disconnected by localized electrons (pins), is stabilized
thanks to the kinetic energy of those rings, as discussed in Sec.
\ref{Variational}.
Except for the semi-metal, all the phases reported in this phase diagram break
the translational symmetry, as a finite $M(K)$ indicates (see Fig.\ref{scan}).
We provide in Sec. \ref{TL} the complete analysis about the nature of these
phases by UHF, and then we describe the SDCI method and apply it to confirm their
stability upon the correlations.
We have followed the following strategy. First, we identify the phases by the
UHF. Then, we check their stability deep in their domains and close to the
phase boundaries by constructing  a configuration interaction effective
Hamiltonian in the single and double excitations (SDCI) Slater basis (see the next
section for more details). We also verify that no new phases arise.
Finally, in Sec. \ref{Variational}, we unravel the mechanism of our
phases by constructing a trial wave function and a perturbative theory based on our UFH
and SDCI results at large $U$ and $V$ limit. This validates the present phase
diagram.
%


\section{Identification of the phases and Thermodynamic Limit}
\label{TL} 
\subsection{UHF thermodynamic limit}
In this section, we first present the thermodynamic limit (TL) phase diagram,
displayed in Fig.\ref{PhaseDiag}, extracted from finite-size scalings of the
UHF results. Then, we confirm the existence and the stability of all the phases
beyond the mean-field level thanks to the single and double configuration
interaction (SDCI).
We have performed finite size scalings on the four order parameters, $M({\bf
q})$, $\xi^{\pm}$, and $\rho^-_A \rho^-_B \rho^-_C$ on clusters up to $3 \times
36 \times 36$ sites. To illustrate our results, we consider the scan line
displayed in Fig.\ref{PhaseDiag}, which passes through all the phases.  The
results are displayed in Fig.\ref{scan}.
As one can see, size effects are important (inset), as expected for the kagome
lattice, but thanks to the large size clusters available, precise size scalings
are performed.
Our analysis reveals two second order phase transitions, between the SCDW and
the SM and between the SM and the CDW. The finite order parameter $M(K)$ (blue)
at the TL indicates the spontaneous symmetry breaking of translations at the {\it K}
point and thus a $\sqrt{3} \times \sqrt{3}$ unit cell of nine sites.
The quantity $\rho^-_A \rho^-_B \rho^-_C$ (gray) is zero as soon as on each
triangle at least one zero magnetization site is present or finite otherwise.
We observe that the PMD phase is the only one with a finite TL value, and all sites
of each triangle are occupied.
Finally, $\xi^{+(-)}$ (red, green) provide insights about the internal charge
(magnetization) structure of each triangle.
In Table~\ref{OrderParam} is reported which order parameter is compatible
with which phase. 
\begin{table}[ht]
\begin{center}
    \begin{tabular}{ | l | c | c | c | c |}
    \hline
Phases &   $ M({\bf K})$ & $\xi^+$ & $\xi^-$ & $\rho^-_A \rho^-_B \rho^-_C$ \\ \hline \hline
SM     &   0    &   0   &   0   &  0    \\ \hline
CDW    &   \checkmark    &   \checkmark   &   0   &  0    \\ \hline
PMD    &   \checkmark    &   \checkmark   &   \checkmark   &  \checkmark    \\ \hline
SCDW   &   \checkmark    &   \checkmark   &   \checkmark   &  0    \\ \hline
    \end{tabular}
\caption{
\label{OrderParam} Classification of the possible phases that may occur in
the extended Hubbard model on the kagome lattice. Such phases can be distinguished
from the density-density correlation $M({\bf K})$, the internal triangular
structures in charge ($\xi^+$) and magnetization ($\xi^-$), and the presence or
not of a $s^z=0$ site per triangle ($\rho^-_A \rho^-_B \rho^-_C$).
}
\end{center}
\end{table}
All of them have been identified unambiguously within the UHF approach as seen
in Fig.\ref{scan}. The corresponding density maps of the PMD and the SCDW,
for the charge and the magnetization, are very close to the ones
obtained by the SDCI (see below) depicted in  Fig.\ref{snapshots}.

\begin{figure}[ht!]
\includegraphics[width=0.165\textwidth,clip]{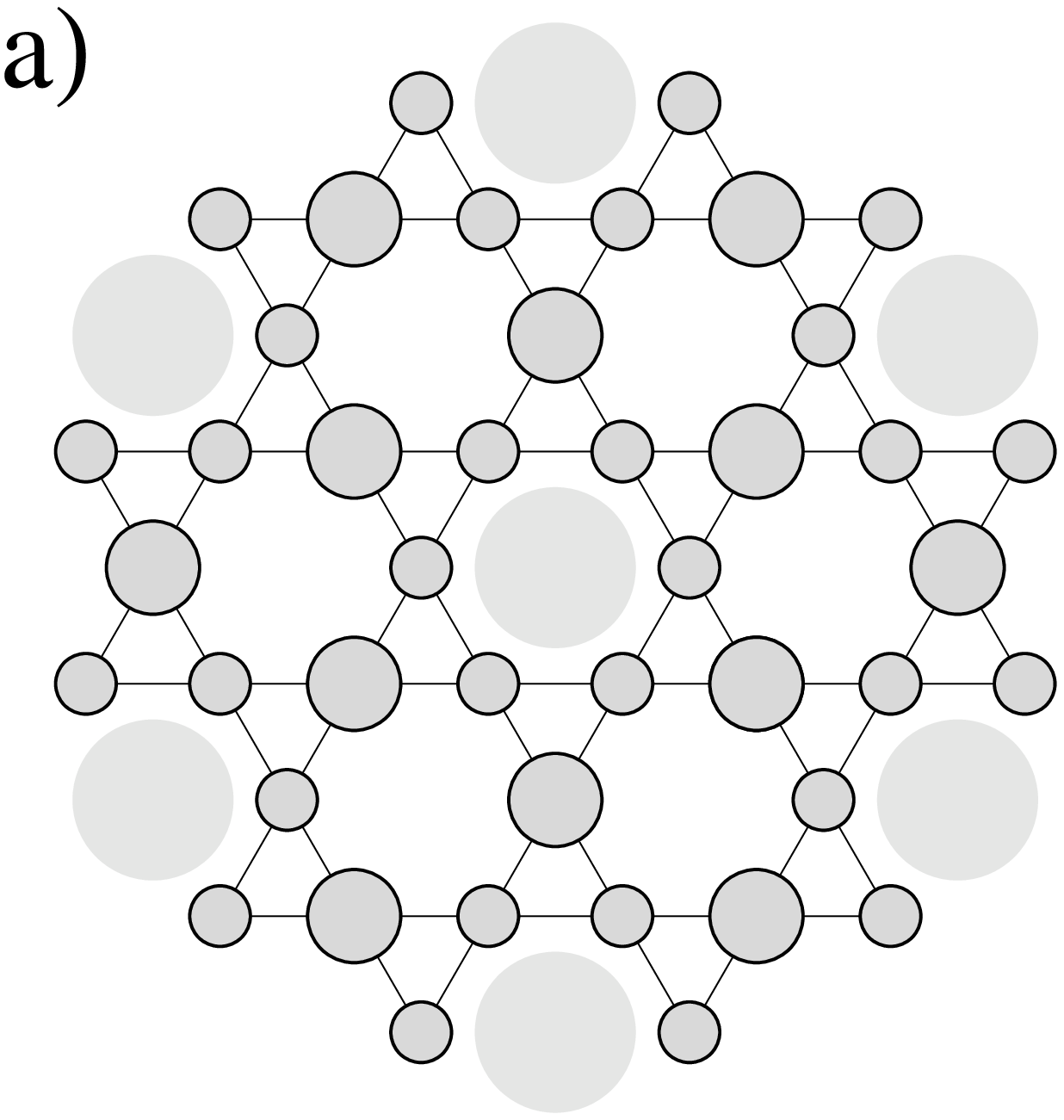}
\includegraphics[width=0.165\textwidth,clip]{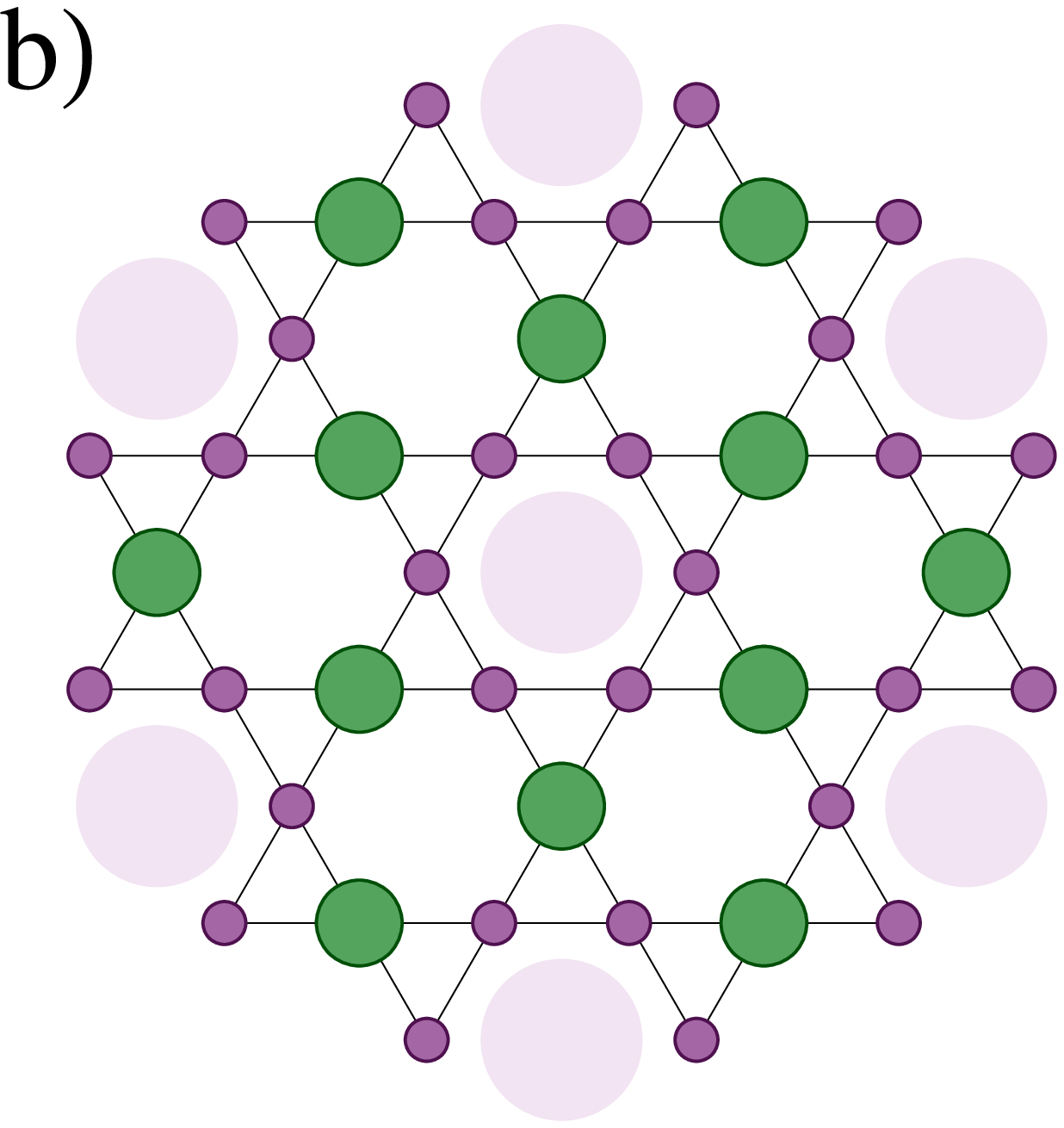}
\\
\includegraphics[width=0.165\textwidth,clip]{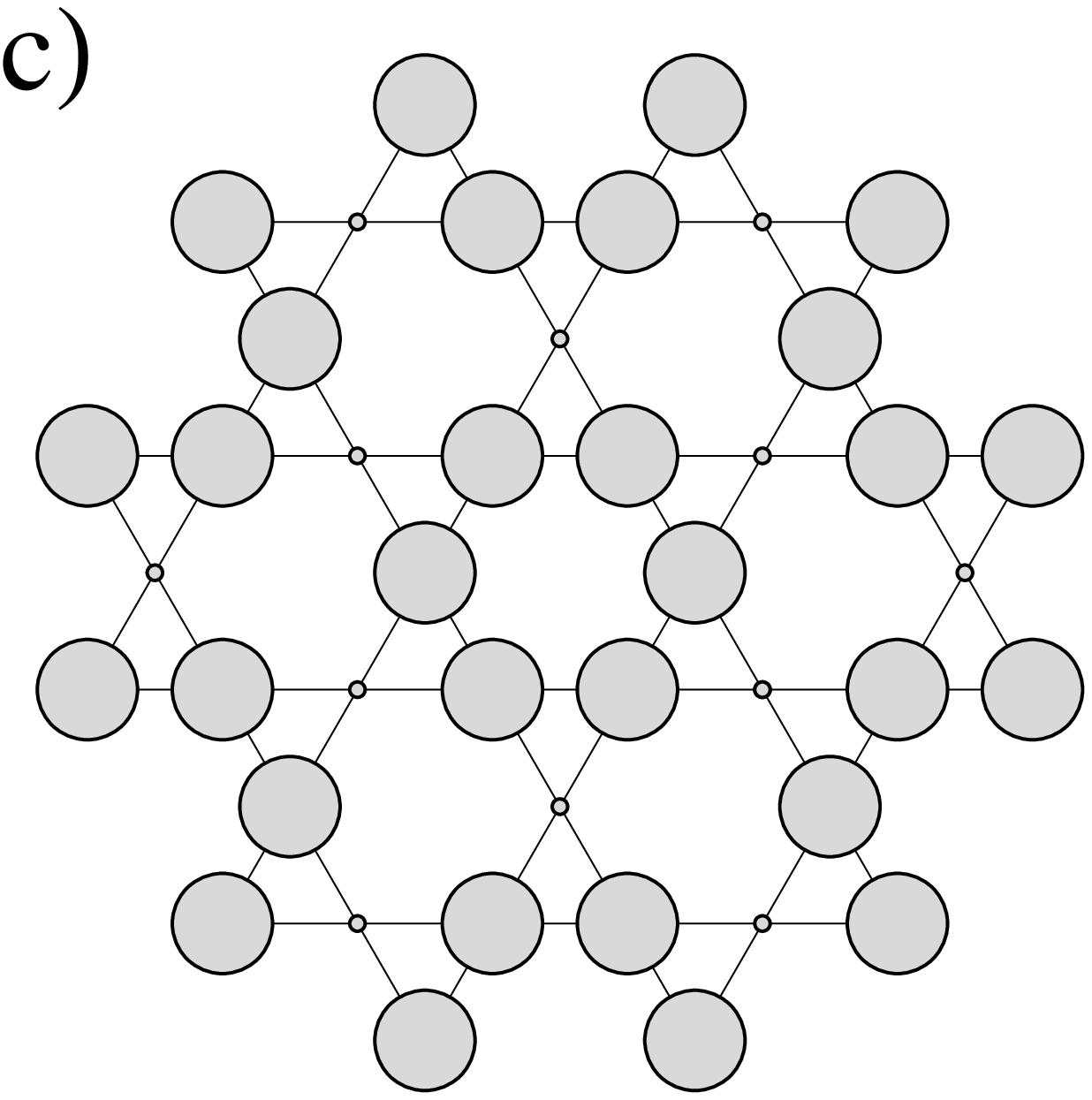}
\includegraphics[width=0.165\textwidth,clip]{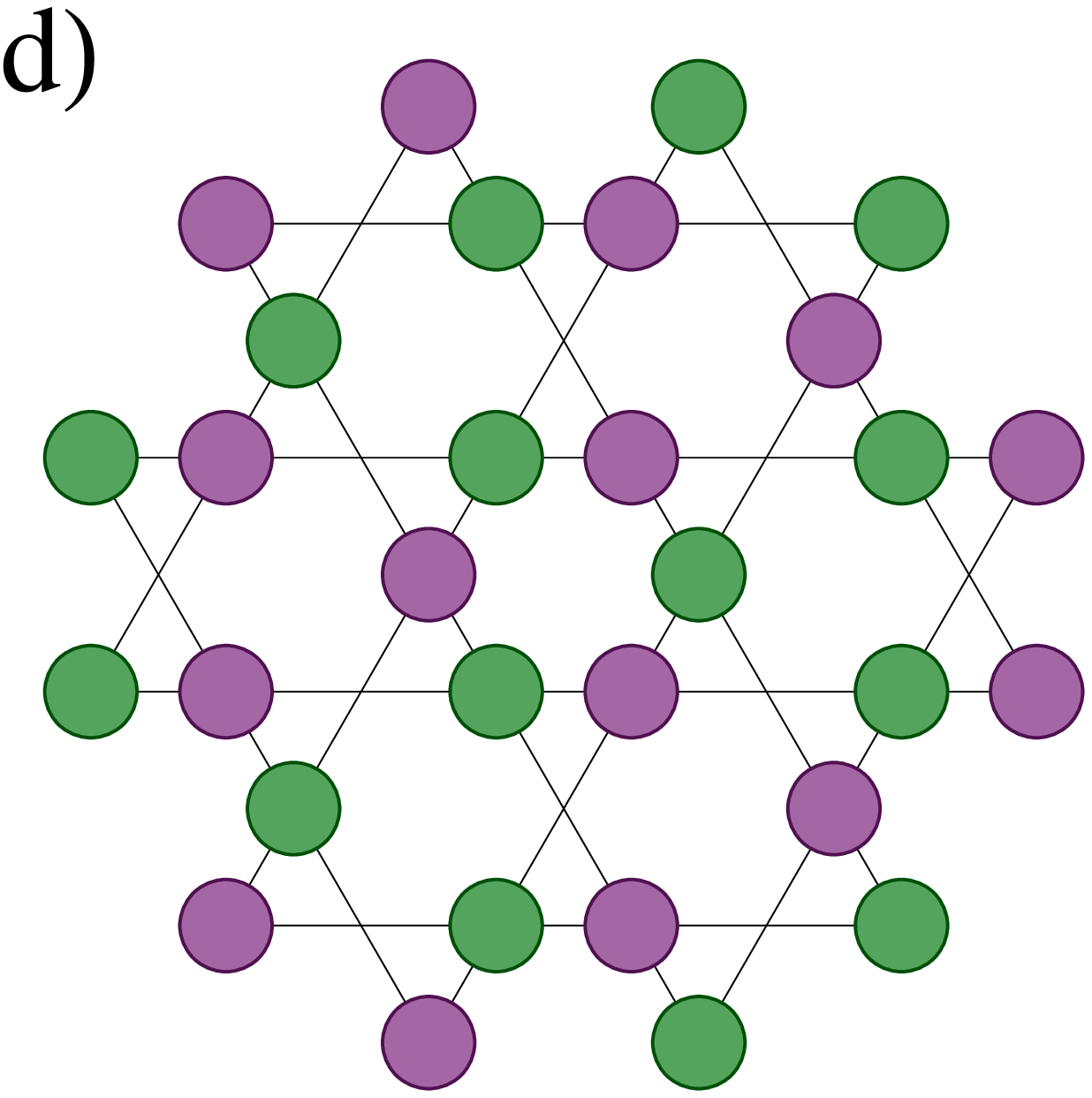}
\caption{(Color online). Snapshots of local charges (a,c) and spins (b,d) for
the PMD phase (a,b) at $U=8$ and $V=0$ and SCDW (c,d) at $U=8$ and $V=2$
obtained by SDCI on a cluster of 108 sites. The larger the circle, the higher the
density (charge and spin). The colors green and purple correspond respectively
to spin sectors up and down.  The larger and lighter disks in (a) and (b)
illustrate the metal droplets standing on the hexagons of the PMD phase, pinned by localized electrons (green circles in (b)).
\label{snapshots}
}
\end{figure}
\subsection{Configuration interaction: beyond the Hartree-Fock}
In order to get further insights about the nature of the phases and their
stability beyond the UHF, we have developed a {\it configuration interaction}
method which aims to bring back the correlations on top of a mean-field
single particle solution, here obtained by the UHF (see Sec.\ref{methods}).
In Table~\ref{tableCI} are reported the ground state energies obtained by ED
(FCI), SDCI and UHF on a $2 \times 2$ cluster, the largest available by ED at
the 1/3 filling. Keeping all single and double excitations above
the Fermi sea corresponds to a total of 3489 Slater determinants. The
parameters are chosen in the three non trivial phases, the
CDW, the SCDW and the PMD. For each of them, the SDCI clearly improves the GS
energies, hence showing that correlations have been incorporated.

In our model, if the phases obtained by UHF are not stable under the
correlations (for example the UHF is known to overestimate ferromagnetism), we
expect the SDCI to stabilize other solutions.
\begin{table}[ht]
\begin{center}
    \begin{tabular}{ | l | r | r | r | r |}
    \hline
Phases &   $(U,V)$ & UHF & SDCI & ED (FCI) \\ \hline \hline
PMD     &   $(8,0)$ & $-10.818(4)$  & $-12.801(5)$  & $-14.071(9)$ \\ \hline
SCDW   &   $(8,2)$ & $ 6.819(1)$  & $ 5.199(0)$  & $ 4.032(1)$  \\ \hline
CDW    &   $(0,2)$ & $-5.820(9)$  & $-5.985(6)$  & $-6.430(1)$  \\ \hline
    \end{tabular}
\caption{\label{tableCI} Ground state energies of the different phases obtained
by the three methods used in this work, here on a $2 \times 2$ cluster (12
sites) for various parameter sets.}
\end{center}
\end{table}
We were able to distinguish the four phases identified at the mean-field level. The
snapshots of the local densities $n_{i,\sigma}$ of the PMD and the SCDW phases
are depicted in  Fig.\ref{snapshots}.
By turning on the correlations on top of the UHF solution, we found that all
the phases are stable up to the largest cluster available by the CI, in our
case the $6 \times 6$ unit cells (108 sites). Note that for such a large cluster,
the corresponding basis contains 263226945 determinants, which is not reachable
by our computers. In order to reduce this size, it is possible to keep among the
single and double excitations only those for which the excitation energy does not
exceed a chosen cutoff $\Lambda_{\textrm{CI}}$. By adjusting correctly this cutoff, one can
find a compromise between the improvement of the GS wave function and the size of the basis we want to deal with. 
For the $3 \times 6 \times 6$ site cluster, we have kept 1850220 Slater
determinants for energy improvement of the order of $10\%$ on the GS energies.
On the $3 \times 3 \times 3$ site cluster, for which no cutoff is required
(97039 Slater determinants), the UHF energy is $14.4466(8)$ and $11.4323(0)$
for the SDCI in the SCDW phase at $U=8$ and $V=2$ [an improvement of
$20.97(5)\%$]. In the PMD phase at  $U=8$ and $V=0$, the UHF gives $-24.9667(2)$
against $-28.7069(6)$ for the SDCI [$14.98(0)\%$].
In Fig.\ref{snapshots} are depicted the local charge and spin densities
obtained by SDCI on the 108 site cluster, for the PMD and the SCDW
phases.  For information, in our computations, the change in the density values
between the SDCI and the UHF never exceeded 3-4\%, even close to the phase
boundaries. This indicates that (i) these phases are stable upon the
correlations, and (ii) the UHF phase diagram at the TL of Fig.\ref{PhaseDiag}
captures the essential physics. 
Let us conclude this section by a remark concerning the phase boundaries of
Fig.\ref{PhaseDiag}. In this work, we have employed the SDCI to validate the
presence of the main phases beyond the UHF and to verify whether or not new
phases could be stabilized. Hence, we have focused deep in their domains, and
made checking close to the main transitions. The question of the exact
location of the boundaries within this approach requires an optimized
methodology in selecting the active determinants, in order to reduce the size
of the basis while keeping the correlations strong\cite{Li,Bunge,Sambataro}. This is left for another
work.
%


\section{Trial wave function at large U and V}
\label{Variational}
Let us now try to understand the characteristics of the PMD and the SCDW phases
by considering the large $U$ and $V$ limit.

\begin{figure}[ht] \includegraphics[width=0.40\textwidth,clip]{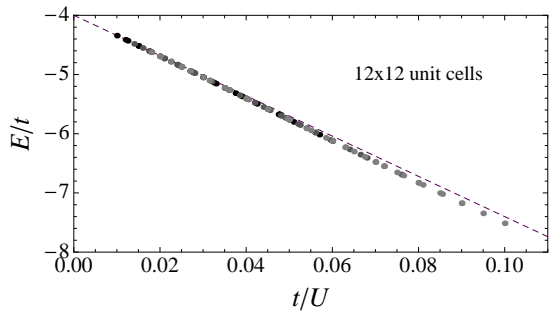}
\caption{Agreement between the UHF energy and the
trial wave function based on metallic hexagons for the PMD phase. We show
several scans from $t=0.5$ (darker points) to $t=1$ (lighter points). Each scan
is made from $U=10$ to $50$. The dashed line is the energy expected in 
the PMD phase $E = -4t -34 \frac{t^2}{U}$. 
\label{PMDmodel}
}
\end{figure}

At large $U \gg V$, we find a phase which is the separation between {\it pins}
(one electron) on each site of an enlarged kagome lattice and {\it balls} of three
polarized electrons on each hexagon, that we expect to be in a local metal
state (delocalized electrons on six sites). Indeed, as a first approximation, one could think that
these hexagons are isolated by the pins, as suggested  by the real space
density maps of the PMD depicted in Figs. \ref{snapshots}(c) and \ref{snapshots}(f).
To validate this picture, we have calculated the energy of the
trial wavefunction (TWF)
$ | \psi \rangle  =
[ 
\underset{i}{\otimes} | i \rangle 
] 
\otimes 
[ 
\underset{h}{\otimes} | \hexagon_h \rangle 
] 
$ where the first product is over single polarized electron states
$\{ | i \rangle \}$ standing on the sites of the enlarged kagome lattice and the
second is over all the remaining hexagons, $|\hexagon_h \rangle$ being the ground state of
an extended Hubbard model of three inversely polarized electrons on a ring:
\begin{eqnarray}
\label{effham}
H_{\hexagon} =  -t  \sum_{i \in \hexagon} \left[ ( c_i^\dagger c_{i+1} +
h.c ) + V n_i n_{i+1} \right].
\end{eqnarray}
By computing the ground state energy and the kinetic operator $ \hat{P} =
(c_0^+ c_1 c_2^+ c_3 c_4^+ c_5 + h.c.)$ (the sites are labeled clockwise from
zero to five), an inflexion point appears about $V \simeq 3t$. For $V < 3t$, the
ground state is delocalized (metal). At $V=0$, the ground state on each hexagon
is a simple metal of three electrons, and in terms of local momenta $k$ on the
six-site ring, the wave function is trivially given by the product state of the
single electron Bloch states $|\hexagon_h \rangle = |k^0 \rangle \otimes |k^{+}
\rangle \otimes |k^{-} \rangle$, where $k^0 = 0$, and $k^{\pm} = \pm \pi / 3$,
with energies $E_{0} = -2 t$ and $E_{\pm} = - t$ respectively. By applying the
perturbation theory on $| \psi \rangle$ at large $U$, we find a correction to
the energy as $E_{k^0} = -2 t -4 \frac{t^2}{U}$, $E_{k^\pm } = - t -3
\frac{t^2}{U}$, and for the pins $E_{i} = -24 \frac{t^2}{U}$, for a total
trial energy $E_{\textrm{TWF}} = -4 t -34 \frac{t^2}{U}$.
In Fig.\ref{PMDmodel} are depicted the GS energies at $V=0$ obtained by UHF and
the trial energy obtained from the TWF with metallic hexagons; their
agreement is excellent.
%

We turn now tn the SCDW. In the large $U$ limit, at $t=0$, there are no double
occupancies and the system, for any finite $V$, enters an {\it ice-rule}
constraint of two electrons per triangle. Its ground state manifold is
macroscopically degenerate and consists of closed loops of various lengths.
When the quantum fluctuations $t$ are turned on, the UHF and the SDCI select
the smallest possible loops as observed in Fig.\ref{snapshots}-d.
From a perturbation theory, there are two types of processes which can lift
this degeneracy: an antiferromagnetic coupling of spins of order $\sim
t^2/(U-V)$ at the second order, and resonating plaquette terms (ring exchange
of three electrons on a hexagon) in $\sim t^3/V^2$ at the third order, similar to
those of the QDM\cite{Nishimoto} at $U=0$ and $V \gg t$.
Unfortunately, neither the UHF (mean-field) nor the SDCI (up to double
excitations) allow us to distinguish which process is dominant in functions of $U$
and $V$. All we can say is that, at large enough $V$ and $U \gg V$, the
Heisenberg mechanism will be obviously favored.
But let us go back to Eq.~(\ref{effham}) and $| \psi \rangle$. The inflexion point
in $ \langle \hat{P} \rangle$ appearing around $V \simeq 3 $
indicates that $\sim t^3/V^2$ ring exchange terms become dominant for higher
$V$ and thus the PMD is destroyed at such large values of $V$. At the mean-field
level, the UHF snapshot would be similar to that of Fig.\ref{snapshots}-d, but
now the resonating hexagons are no longer the same as the Heisenberg loop 6
discussed above.
Hence, two scenarios are possible at the transition of the PMD: (i) the
system enters a phase with resonating plaquettes separated by pins
based on our trial wave function, and then a further transition to a spin state
driven by Heisenberg couplings is stabilized (with no pins), or (ii) the
Heisenberg state is selected right away.
Our UHF and SDCI results present a phase transition about $V \simeq 0.25$ up to
$U = 12 t$, which is much smaller than $V \simeq 3$, at which the internal
structure of the hexagon is changing. This seems to indicate that scenario (ii)
is favored,
or in this case, based on the results of Fig.\ref{snapshots}-d, a ground state
made of six-spin singlet hexagons of energy per site $E_6 \simeq -0.467(1)$ is
selected among the degenerate configurations. This is what we call the
spin charge density wave (SCDW).
Finally, let us compare our results with previous work. In Wen {\it et al.}
\onlinecite{Wen}, since no translational symmetry breaking is allowed, a stripe
phase with disconnected antiferromagnetic chains is obtained. In this case,
the quantum fluctuations of the perturbation theory lowers the classical energy
by $E_c = \frac{1}{4} - \log 2 \simeq -0.443(1)$. The hexagonal spin chain is
then more likely when the symmetry constraint is removed, as confirmed by the
snapshots of the SCDW depicted in Fig. \ref{snapshots}(c) and \ref{snapshots}(d).

\section{Concluding remarks} 
In this paper, we have studied an extended Hubbard model of spinful electrons
on the kagome lattice at 1/3 filling.
In addition to two already known phases, the semi-metal (SM) and the charge
density wave (CDW), we have established the presence of two original phases,
the pinned metal droplets (PMDs) and the spin charge density wave (SCDW) stabilized at
intermediate $U$ and $V$.

We have first used an unrestricted Hartree-Fock approach in order to extract
the phase diagram at the TL.
We have then constructed an effective configuration interaction Hamiltonian by
projecting the microscopic model onto a basis of Slater determinants containing
the UHF wave function, and all single and double excitations (SDCI).  This
allows for bringing  back partial correlations on top of the UHF solution.
For various sets of parameters, the ground-state energy (and related GS) is
strongly improved (up to 60\% of the correlation energy).  We have confirmed that (i) no new phases are
stabilized by the correlations and (ii) the two original phases, namely the PMD
and the SCDW, are very robust against the correlations.

Finally, to get more insights about the nature of two new phases, we have
performed a perturbation theory on a trial wave function. We have
shown that the properties of the pinned metal droplets (PMDs) are perfectly captured by
the UHF and explained by our trial wave function.  It consists of
polarized pins (electrons) standing on the sites of an enlarged kagome lattice,
separating droplets of metal living on the remaining hexagons.
The second important phase, the spin charge density wave (SCDW), is driven by
Heisenberg interactions and leads to loop-6 ring isolated spin states.
Note that we have discussed the possibility of a third phase that could arise
in the system, close to the PMD but with resonating plaquettes instead of
metal droplets on the remaining hexagons. In some sense, this would
correspond to an internal crossover of the hexagon states, the pins remaining
unchanged.
However, within our approaches, this phase does not seem to be stabilized at
the intermediate parameters considered in this work. Instead, the PMD and the
six-spin singlet SCDW are the more stable.

We recently became aware of
related work by F. Pollmann {\it et. al} (Ref.~[\onlinecite{Pollmann2}]), where
the authors have considered an effective model  of Eq.~(\ref{ham}) at large $U$
and $V$ limit. We thank F. Pollmann for stimulating discussions about their
results corroborating the existence of the phases reported in this manuscript.

{\it Acknowledgements} -- A.R. is grateful to F. Mila for insightful remarks
about the stability of the phases. We thank S. Fratini for enlightening
discussions and his careful rereading of the manuscript. A.R. thanks G.
Bouzerar and M.-B. Lepetit for their valuable comments about the methods used
in this paper.
This work is supported by the French National Research Agency through Grants
No. ANR-2010-BLANC-0406-0 NQPTP and No. ANR-12-JS04-0003-01 SUBRISSYME.

\end{document}